\documentstyle[12pt]{article}
\newcommand{\be}{\begin{equation}}
\newcommand{\ee}{\end{equation}}
\newcommand{\bea}{\begin{eqnarray}}
\newcommand{\eea}{\end{eqnarray}}
\newcommand{\nn}{\nonumber}
\newcommand{\p}{\phi}
\newcommand{\q}{\Omega}
\newcommand{\hp}{\hat{\phi}}
\newcommand{\hq}{\hat{\Omega}}
\newcommand{\PP}{\hat{P}}

\newcommand{\F}{\cal{F}}
\newcommand{\M}{{\cal{M}}}

\begin{document}
\title{Abelianization of First Class Constraints}
\author{F. Loran\thanks{e-mail:
loran@cc.iut.ac.ir}\\ \\
  {\it Department of  Physics, Isfahan University of Technology (IUT)}\\
{\it Isfahan,  Iran,} \\{\it and}\\
  {\it Institute for Studies in Theoretical Physics and Mathematics (IPM)}\\
{\it P. O. Box: 19395-5531, Tehran, Iran.}}
\date{}
\maketitle

\begin{abstract}
We show that a given set of first class constraints becomes abelian if one maps each
constraint to the surface of other constraints. There is no assumption that first
class constraints satisfy a closed algebra. The explicit form of the projection map
is obtained at least for irreducible first class constraints. Using this map we give
a method to obtain gauge fixing conditions such that the set of abelian first class
constraints and gauge fixing conditions satisfy the symplectic algebra.
\end{abstract}

\newpage
\section{Introduction}
It is known that first class constraints are responsible for the appearance of gauge
freedom \cite{Dirac}. Given a set of first class constraints, the generator of gauge
transformation is a nontrivial combination of first class constraints that satisfies
some conditions derived in reference \cite{Pons}. This combination is the simplest if
first class constraints are abelian i.e. when the Poisson brackets of these
constraints with each other vanish identically.
 \par
 Abelianization of first class constraints has two
more important consequences. First, following Dirac's arguments, quantization of a
set of first class constraints $\p_i$'s satisfying a closed algebra,
 \be
 \{\p_i,\p_j\}=C_{ijk}\p_k,
 \label{int1}\ee
 in which the structure functions $C_{ijk}$'s are not $c$-numbers requires a definite
 operator ordering \cite{Dirac}. That is because in Dirac quantization, physical states are
 defined as null eigenstates of the operators $\hp_i$'s,
 \be
 \hp_i\left|\mbox{phys}\right>=0,
 \label{int2}\ee
 in which the operator $\hp_i$'s are defined corresponding to the constraints
 $\p_i$'s. Definition (\ref{int2}) and the algebra (\ref{int1}) are
 consistent if the operators $\hat{C}_{ijk}$'s, defined corresponding to the structure
 functions $C_{ijk}$'s, sit on the left of the operators $\hp_i$'s similar to
 Eq.(\ref{int1}). The existence of such an operator ordering is not evident generally.
 Apparently, when first class constraints are abelian, no such operator ordering
 should be considered. Second, in BRST formalism, the algebra (\ref{int1}), in
 general, leads to a complicated expansion of the BRST charge in terms of the ghosts.
 When first class constraints are abelian, the generator of BRST transformation can be
 recognized in the most simple way. For details see \cite{Hen}.
 \par
 Abelianization procedure
can be realized via constraint resolution \cite{Hen,Gold} or via generalized canonical
transformation for general non-abelian constraints (that satisfy a closed algebra)
\cite{Berg}. In reference \cite{Gog} the authors study abelianization via Dirac's
transformation. In this method, one assumes that linear combinations of non-abelian
first class constraints (satisfying a closed algebra) exist that converts the given
set of non-abelian constraints to an equivalent set of abelian constraints. In this
way the problem of abelianization is led to that of solving a certain system of first
order linear differential equations for the coefficients of these linear combinations.
 \par
 In present work, in section 2, we show that, in general, if one maps each first
class constraint to the surface of the other constraints, the resulting first class
constraints are abelian. The advantageous of this result is that since the explicit
form of the above mentioned map can be obtained, one can {\it practically} and {\it
globally} make an arbitrary set of first class constraints abelian. Of course our
method is applicable only when constraints belong to $\F$, the set of "well behaved
functions" of phase space coordinates. Here, by "well behaved function" we mean those
functions that satisfy the validity conditions of Cauchy Kowalevski theorem
\cite{John}. Section 3 is devoted to conclusion and some discussions.
\section{The Method of Abelianization}
Cauchy Kowalevski theorem implies that for a given function $\p\in\F$, the partial
differential equation,
 \be
 \{\p,\q\}=\sum_i a^\mu\frac{\partial}{\partial z^\mu}\q=1,
 \label{a1}
 \ee
 in which $a^\mu=\{\p,z^\mu\}$ and $z^\mu$'s are the phase space coordinates,
 has at least one solution. To obtain $\q$ conjugate to a
 given $\p$, i.e. $\{\p,\q\}=1$, it is sufficient to find a function $H\in\F$,
 such that $\hp^{M+1}H=0$ for an integer $M\geq1$, where the operator $\hp$ is defined
 via the relation
 \be
 \hp\xi\equiv\{\p,\xi\},\hspace{1cm}\xi\in\F.
 \label{def1}\ee
 One can show that the coefficient of $\hp^M H$ in $\hp^{M-1}H$ is a solution of
 Eq.(\ref{a1}) as follows \cite{symplectic}.\\
 Since $\q$ exists and satisfies Eq.(\ref{a1}), there is a local coordinate transformation
 $z^\mu\to Z^{\mu'},\q,\p$, and the function $H(z^\mu)=H\left(z^\mu(\q,\p,Z^{\mu'})\right)$
 can be written as a polynomial in $\q$,
 \bea
  \label{k1}
  H\left(z^\mu(\q,\p,Z^{\mu'})\right)&=&\sum_{m=0} \frac{A_m(0,\p,Z^{\mu'})}{m!}\q^m,\\
  \label{k2}
  \hp A_m &=& \frac{\partial}{\partial \q}A_m(0,\p,Z^{\mu'})=0,\hspace{2cm}m\geq0.
  \eea
  Comparing $\hp^M H=A_M$ with $\hp^{M-1}H=A_M\q+A_{M-1}$, one can determine $\q$ as
 the coefficient of $\hp^M H$ in $\hp^{M-1}H$. It should be noted that if
 $A_{M-1}=\tilde{A}A_M+A_{M-1}^0$ then $\q$ obtained in this way becomes equal to
 the assumed $\q$ ($\q_s$) plus $\tilde{A}$. Since $\hp\tilde{A}=0$, the obtained
 $\q$, though is not equal to $\q_s$, but satisfies Eq.(\ref{a1}).
 \par
 Let's define the operator $\PP_\p$,
 \be
 \PP_\p\equiv e^{\p\hq}=\sum_{n=0}\frac{1}{n!}\p^n\hq^n.
 \label{a3}
 \ee
 which satisfies the property,
 \be
 \hq\PP_\p=0,
 \label{a2}
 \ee
 The validity of Eq.(\ref{a2}) can be directly checked using the general properties of Poisson
 brackets as follows.
 \bea
 \hq\PP_\p&=&\hq\sum_{n=0}\frac{1}{n!}\p^n\hq^n\nn\\
 &=&\sum_{n=0}\frac{1}{n!}\left(n\{\q,\p\}\p^{n-1}\hq^n+\p^n\hq^{n+1}\right)\nn\\
 &=&0
 \eea
 where in the last equality we have used $\{\q,\p\}=-1$. \\
 {\bf Lemma 1.} The operator $\PP_\p$ is the projection map to the subspace of phase
 space defined by $\p=0$, i.e. for any $\xi\in\F$, $\PP_\p \xi=\xi|_{\p}$.
 \par
 {\bf Proof}. From Eq.(\ref{a2}) and definition (\ref{a3}), one verifies that
 $\PP_\p^2=\PP_\p$, i.e. $\PP_\p$ is a projection map. Since $\p$ and $\q$ are
 conjugate ($\{\p,\q\}=1$), there exist a (local) coordinate transformation
 $z^\mu\to Z^{\mu'},\p,\q$, such that $\xi(z)=\tilde{\xi}(\p,\q,Z)$ and
 $\hq=-\frac{\partial}{\partial \p}=-\partial_\p$. Consequently,
 \be
 \PP_\p \xi(z)=e^{-\p\partial_\p}\tilde{\xi}(\p,\q,Z)=\tilde{\xi}(0,\q,Z)=\xi|_{\p}.
 \ee

 {\bf Lemma 2.} Considering a function $\xi\in\F$ and a conjugate pair of functions
 $\p$ and $\q$, we have
 \be
 \xi=\xi|_{\p}\hspace{5mm}\mbox{iff}\hspace{5mm}\hq \xi=0.
 \label{a5}
 \ee
 {\bf Proof.}
 \par
 $a)$ If $\xi=\xi|_{\p}$ then from lemma 1, $\xi=\PP_\p \xi$. Therefore using Eq.(\ref{a2}),
 $\hq \xi=\hq \PP_\p \xi=0$.
 \par
 $b)$ if $\hq \xi=0$ then $\xi=\PP_\p \xi=\xi|_{\p}$.
 \\
 {\bf Corollary 1.}
 \be
 \{\xi|_{\p},\zeta|_{\p}\}= \{\xi|_{\p},\zeta|_{\p}\}_\p.
 \label{a6}
 \ee
{\bf Corollary 2.}
 \be
 \{\xi|_{\p},\p\}= \{\xi|_{\p},\p\}_\p.
 \label{a7}
 \ee
 These corollaries can be proved using the Jaccobi identity to show that the Poisson
 brackets of the LHS of Eqs.(\ref{a6},\ref{a7}) with $\q$ is vanishing.\\
 {\bf Lemma 3.} If $\p=\p|_{\psi}$ then $\psi=\psi|_{\p}$.
 \\
 {\bf Proof.} Since there exist a function $\q$ conjugate to $\p$, using the operator
 $\PP_\p$, one can write $\psi$ as a polynomial in $\p$ (similar to Eqs.(\ref{k1},\ref{k2})),
 \be
 \psi=\sum_{i=1}a_i\p^i+\psi|_{\p}.
 \ee
 where $\hq a_i=0$, $i\geq1$. If $a_i$'s do not vanish, the assumption $\p=\p|_{\psi}$
 implies that $\psi(\p)=0$. Thus if $\psi\neq0$, then $a_i$'s should vanish and
 $\psi=\psi|_{\p}$.\\
{\bf Lemma 4.} If
 \be
 \{\p_1,\q_1\}=\{\p_2,\q_2\}=1,
 \ee
 and
 \bea
 \label{la1}
 \p_2=\p_2|_{\p_1},\\
 \label{la2}
 \q_2=\q_2|_{\p_1},
 \eea
 then the operators $\hat{P}_1$ and $\hat{P}_2$ defined corresponding to the functions
 $\p_1$ and $\p_2$ (see Eq.(\ref{a3})) commute with each other. \\
 {\bf Proof.} It is sufficient to show that the operators $\p_1\hq_1$ and $\p_2\hq_2$
 commute i.e.
 \be
 [\p_1\hq_1,\p_2\hq_2]=0.
 \ee
 First we show that
 \be
 [\hq_1,\p_2\hq_2]=0.
 \ee
 Considering a function $\xi\in\F$, one verifies that,
 \bea
 \left(\hq_1(\p_2\hq_2)\right)\xi&=&\left\{\q_1,\{\p_2\{\q_2,\xi\}\right\}\nn\\
 &=&\p_2\{\q_1,\{\q_2,\xi\}\}\nn\\
 &=&\p_2\{\q_2,\{\q_1,\xi\}\}\nn\\
 &=&(\p_2\hq_2)\hq_1\xi.
 \eea
 The first equality is the result of definition (\ref{def1}). Second equality is
 obtained because $\{\p_2,\q_1\}=0$ as a result of Eq.(\ref{la1}) and lemma 2.
 Third equality can be derived using the Jaccobi identity and noting that from lemma 2 and
 Eq.(\ref{la2}) we have $\{\q_1,\q_2\}=0$.\par
 From lemma 3 and Eq.(\ref{la1}) one verifies that
 \be
 \p_1=\p_1|_{\p_2},
 \ee
 thus $\{\p_1,\q_2\}=0$ which, using the Jaccobi identity, gives,
 \be
 [\p_1,\hq_2]=0.
 \ee
 {\bf lemma 5.} Given a set of first class constraints $\p_i$'s satisfying the algebra
 (\ref{int1}), there exist a function $\q_i$ that satisfies the following properties:
 \bea
 \label{lm42}
 \{\p_i,\q_i\}&=&1,\\
 \label{lm43}
 \q_i&=&\q_i|_{\M},\hspace{1cm} .
 \eea
 where by $\M$ here, we mean the surface of all constraints $\p_i$'s.
 \\
 {\bf Proof.} The function $\q_i$ with the above properties can be determined if one
 finds a function $H=H|_{\M}$ such that $\hp_i H\neq0$ but $\hp^{M+1}_i H=0$
 for some integer $M$. See the arguments below Eq.(\ref{def1}) and note that since
 $H=H|_{\M}$ the function $\q_i$ obtained in this way satisfies Eq.(\ref{lm43}).
 Of course one should
 show that there exist a function $H=H|_{\M}$ such that $\{\p_i,H\}\neq 0$.
 To prove this assertion, assume that there does not exist any function with this
 property. Since there exist a function $\q_i$ that satisfies Eq.(\ref{lm42}), this
 assumption leads to the following equality,
 \be
 \left\{\p_i,(\q_i-\q_i|_{\M})\right\}=1,
 \label{lm44}\ee
 because
 \be
 \{\p_i,\q_i|_{\M}\}=0.
  \ee
 In addition,
 \be
 \q_i-\q_i|_{\M}=\sum_{i}a_i\p_i+{\it O}(\p^2),
 \ee
 in which the first term on the RHS is linear in $\p_i$'s i.e. $a_i=a_i|_{\p_j}$ and
 the second term includes all nonlinear terms in $\p_i$'s. Consequently
 Eq.(\ref{lm44}) implies that
 \be
 \{\p_i,(\sum_ja_j\p_j)\}=1.
 \ee
This is not consistent with the algebra (\ref{int1}) because the coefficients $a_i$'s
do not vanish on the surface of constraints and consequently the algebra (\ref{int1})
can not be closed.
 \par
The above results provide a simple method for abelianization of first class
constraints $\p^0_i$, $i=1,\cdots,n_0$ satisfying the closed algebra
 \be
 \{\p^0_i,\p^0_j\}=\sum_k C_{ij}^k\p^0_k.
 \label{a8}
 \ee
 To obtain this result, consider one of the constraints, e.g. $\p^0_1$ and its
 conjugate $\q^0_1$. The operator $\PP_0=e^{\p_1^0\hq^0_1}$ maps all the remaining
 constraints to the constraint surface $\p_1^0=0$. Let's call these mapped constraints
 $\p^1_i$, $i=1,\cdots,n_1\leq n_0-1$. Since $\p^1_i=\p^1_i|_{\p^0_1}$, using lemma 3,
 one realizes that $\p^0_1=\p^0_1|_{\M^1}$, where $\M^1$ is the constraint surface determined
 by the constraints $\p^1_i$'s.
 Considering $\p^1_1$ and its conjugate $\q^1_1$, the
 operator $\PP_1=e^{\p^1_1\hq^1_1}$ maps the remaining constraints to the constraint
 surface $\p^1_1=0$. One should note that the function $\q^1_1$ should be chosen such that
 $\q^1_1=\q^1_1|_{\p^0_1}$, because the map $\hat{P}_1$ should leave the constraints
 on the surface of $\p^0_1$. Lemma 5 proofs the existence of such a function
 $\q^1_1$. Continuing this process, one finally ends up
 with a set of  irreducible constraints $\p^0_1,\p^1_1,\cdots,\p^m_1$, $m\leq n_0$, equivalent
 to the  constraints $\p^0_i$'s, $i=1,\cdots,n_0$. By equivalence, we mean that the constraint
 surface $\M$ defined by the constraints $\p^0_i$'s is equivalent to that determined
 by the constraints $\p^a_1$'s, $a=0,\cdots,m$. We say the set of $\p^0_i$ are equivalent to
 $\p^a_1$'s because $\p^a_1$'s are, by construction, some (nonlinear)
 combinations of $\p^0_i$'s.\\
 {\bf Lemma 6}. The set of constraints $\p^a_1$'s, satisfy the following relation,
 \be
 \p^a_1=\p^a_1|_{\p^b_1},\hspace{1cm}b\neq a.
 \label{cal1}\ee
 {\bf Proof}.  First we show that
 \be
 \p^a_1=\p^a_1|_{\p^b_1},\hspace{1cm}b<a,\ a=1,\cdots,m.
 \label{call2}\ee
 For this it is sufficient to prove that
 \be
 \{\q^b_1,\p^a_1\}=0,\hspace{1cm} b<a,
 \label{call3}\ee
 then the validity of Eq.(\ref{call2}) can be verified using lemma 2.
 To prove Eq.(\ref{call3}) we note that: \par
 1) by construction,
 \be
 \p^a_1=\hat{P}_{a-1}\cdots\hat{P}_0\p^a_1
 \label{lem0}\ee
 in which
 \be
 \hat{P}_b=e^{\p^b_1\hq^b_1},
 \ee
 where
 \bea
 \label{lo1}
 &&\p^b_1=\p^b_1|_{\p^c_1},\hspace{1cm} c<b,\\
 \label{lo2}
 &&\q^b_1=\q^b_1|_{\p^c_1}.\\
 \label{lo3}
 &&\{\p^b_1,\q^b_1\}=1.
 \eea
 \par
 Lemma 5 guarantees the existence of $\q^b_1$ satisfying Eqs.(\ref{lo2},\ref{lo3}).
 \par
 2) From lemma 4 and Eqs.(\ref{lo1}-\ref{lo3}) we find that the operators $\hat{P}_{b<a}$
 appearing in Eq.(\ref{lem0}) commute. Eq.(\ref{a2}) implies the validity of
 Eq.(\ref{call3}).
 Using lemma 3 and Eq.(\ref{call2}) one verifies that
 \be
 \p^a_1=\p^a_1|_{\p^b_1},\hspace{1cm}b>a, a=0,\cdots,m-1,
 \label{call5}\ee
which completes the proof.\\
 {\bf Lemma 7.} The set of constraints $\p^0_1,\cdots,\p^m_1$ are abelian.
 \\
 {\bf Proof.} Assume the first class constraints $\p^a_1$'s satisfy a closed algebra
 \be
 \{\p^a_1,\p^b_1\}=\sum_c C^{ab}_c\p^c_1,
 \ee
 for some functions $C^{ab}_c\in\F$. From Eq.(\ref{a6}), and the fact that
 $\p^a_1=\p^a_1|_{\M^{ab}}$ and  $\p^b_1=\p^b_1|_{\M^{ab}}$ in which $\M^{ab}$ is the
 constraint surface determined by all constraints except $\p^a_1$ and $\p^b_1$
 (see lemma 6), we have
 \be
 C^{ab}_c=0\hspace{1cm}c\neq a,b.
 \ee
 In addition $\p^a_1=\p^a_1|_{\p^b_1}$ implies that $C^{ab}_b=0$. Similar argument
 shows that $C^{ab}_a=0$. Therefore
 \be
 \{\p^a_1,\p^b_1\}=0.
 \ee
 This result is interesting because the map between the set of (reducible) non-abelian
 constraints $\p^0_i$'s to the irreducible abelian constraints $\p^a_1$'s is
 globally defined and can be used practically. One should only choose a
 proper constraint at each level, where the projection operator is defined and
 applied. By proper we mean that, for example at first level, $\p^0_1$ should not be a
 multiple of some other constraints. These difficulties disappear if first class
 constraints are generated by the chain by chain method \cite{Self} and consequently
 they are irreducible.
\\
{\bf Lemma 8.} If $\p=\p|_{\psi}$, $\{\p,\psi\}=0$ and $\{\p,\q\}=1$, then
$\{\p,\q|_{\psi}\}=1$.\\
{\bf Proof.} Writing $\q$ as a polynomial in $\psi$,
 \be
 \q=\sum_{i=1}a_i\psi^i+\q|_{\psi},
 \ee
 one verifies that,
 \be
 1=\{\p,\q\}=\sum_{i=1}\{\p,a_i\}\psi^i+\{\p,\q|_{\psi}\}.
 \ee
 Thus,
 \be
 1=\{\p,\q|_{\psi}\}_{\psi}=\{\p,\q|_{\psi}\}_\psi=\{\p,\q|_{\psi}\},
 \ee
 where in the third equality we have used Eq.(\ref{a6}).
 \\
 {\bf Corollary 3.}
 \be
 \q=\q|_{\p}.
 \ee
 {\bf Corollary 4.} The functions $\q^a_1$'s conjugate to $\p^a_1$'s,
 $\{\p^a_1,\q^a_1\}=1$, satisfy the following properties,
 \bea
 \label{b1}
 \{\p^a_1,\q^b_1|_{\M}\}=\delta^{ab},\\
 \label{b2}
 \{\q^a|_{\M},\q^b|_{\M}\}=0.
 \eea
 To prove this equalities use corollary 3, Eq.(\ref{a5}) and note that
 $\p^a_1=\p^a_1|_{\p^{b\neq a}_1}$. The proof is similar to the proof of lemma 7.
 This result is interesting since from
 Eq.(\ref{b1}), one verifies that $\q^a_1|_{\M}$ can be considered as gauge fixing
 conditions, because the equation $\q^a_1|_{\M}=0$
 eliminates the gauge freedom generated by the first class constraint $\p^a_1$'s.
 In this way we have found first class constraints and gauge fixing conditions
 that satisfy symplectic algebra.
\section{Summary}
We showed that irreducible first class constraints obtained by mapping each element
of a given set of first class constraints to the surface of the other elements, are
abelian. The explicit form of these maps at least when constraints are irreducible are
obtained. Final result only depends on general properties of Poisson bracket and do
not depend on its explicit form. Consequently, if system possesses both first class
and second class constraints and first class constraints do not obey a closed
algebra, one can replace ordinary Poisson brackets with Dirac brackets corresponding
to the second class constraints and apply the method studied here to make first class
constraints abelian with respect to Dirac bracket. Furthermore, using Eq.(\ref{a6})
(corollary 1) one can show that if one maps the first class constraints to the
surface of second class constraints, then they satisfy a closed algebra and
abelianization with respect to ordinary Poisson bracket becomes straightforward. The
explicit form of projection map to the surface of second class constraints is studied
in \cite{symplectic}.

\end{document}